\documentclass[preprint,showpacs,preprintnumbers,amsmath,amssymb]{revtex4}
\usepackage{epsfig}
\usepackage{graphicx}
\usepackage{dcolumn}
\usepackage{bm}
\usepackage{threeparttable}

\def\beq{\begin{equation}}
\def\eeq{\end{equation}}
\def\eeqn{\end{equation}}
\newcommand\iden{\leavevmode\hbox{\small1\normalsize\kern-.33em1}}

\newcommand{\bea} {\begin{eqnarray}}
\newcommand{\eea} {\end{eqnarray}}

\let\jnfont=\rm
\def\NPB#1,{{\jnfont Nucl.\ Phys.\ B }{\bf #1},}
\def\PLB#1,{{\jnfont Phys.\ Lett.\ B }{\bf #1},}
\def\EPJC#1,{{\jnfont Eur.\ Phys.\ Jour.\ C }{\bf #1},}
\def\PRD#1,{{\jnfont Phys.\ Rev.\ D }{\bf #1},}
\def\PRL#1,{{\jnfont Phys.\ Rev.\ Lett.\ }{\bf #1},}
\def\MPLA#1,{{\jnfont Mod.\ Phys.\ Lett.\ A }{\bf #1},}
\def\JPG#1,{{\jnfont J.\ Phys.\ G }{\bf #1},}
\def\CTP#1,{{\jnfont Commun.\ Theor.\ Phys.\ }{\bf #1},}
\def\JHEP#1,{{\jnfont JHEP \ }{\bf #1},}
\def\NPPS#1,{{\jnfont Nucl.\ Phys.\ Proc.\ Suppl.\ }{\bf #1},}
\def\CPC#1,{{\jnfont Comput.\ Phys.\ Commun.\ }{\bf #1},}
\def\CPL#1,{{\jnfont Chin.\ Phys.\ Lett. }{\bf #1},}
\def\APPB#1,{{\jnfont Acta\ Phys.\ Polon.\ B }{\bf #1},}
\def\PR#1,{{\jnfont Phys.\ Rept.\  }{\bf #1},}
\def\CHC#1,{{\jnfont Chin.\ Phys.\ C }{\bf #1},}

\def\lsim{\raise0.3ex\hbox{$<$\kern-0.75em\raise-1.1ex\hbox{$\sim$}}}
\def\gsim{\raise0.3ex\hbox{$>$\kern-0.75em\raise-1.1ex\hbox{$\sim$}}}

\begin{document}

\title{\ \\[10mm] A light pseudoscalar of 2HDM confronted with muon g-2 and experimental constraints}

\author{Lei Wang, Xiao-Fang Han}

\affiliation{ Department of Physics, Yantai University, Yantai
264005, PR China \vspace{0.5cm} }


\begin{abstract}
A light pseudoscalar of the lepton-specific 2HDM can enhance the
muon g-2, but suffer from various constraints easily, such as the
125.5 GeV Higgs signals, non-observation of additional Higgs at the
collider and even $B_s\to \mu^+\mu^-$. In this paper, we take the
light CP-even Higgs as the 125.5 GeV Higgs, and examine the
implications of those observables on a pseudoscalar with the mass
below the half of 125.5 GeV. Also the other relevant theoretical and
experimental constraints are considered. We find that the
pseudoscalar can be allowed to be as low as 10 GeV, but the
corresponding $\tan\beta$, $\sin(\beta-\alpha)$ and the mass of
charged Higgs are strongly constrained. In addition, the surviving
samples favor the wrong-sign Yukawa coupling region, namely that the
125.5 GeV Higgs couplings to leptons have opposite sign to the
couplings to gauge bosons and quarks.

\end{abstract}

\maketitle

\section{Introduction}
The ATLAS and CMS Collaborations found a 125.5 GeV Higgs boson at
the LHC \cite{cmsh,atlh}. The latest experimental data show that the
properties of this particle agree with the Standard Model (SM)
predictions. Especially the diphoton signal strength is changed from
$1.6\pm0.4$ to $1.17\pm0.27$ for ATLAS \cite{k-7} and from
$0.78^{+0.28}_{-0.16}$ to $1.12^{+0.37}_{-0.32}$ for CMS \cite{k-8},
which are well consistent with the SM prediction within 1$\sigma$
range. Thus, the 125.5 GeV Higgs signal data can give the strong
constraints on the effects of new physics.

The two-Higgs-doublet model (2HDM) has very rich Higgs
phenomenology, including two neutral CP-even Higgs bosons $h$ and
$H$, one neutral pseudoscalar $A$, and two charged Higgs $H^{\pm}$.
The recent Higgs data have been used to constrain the 2HDM, see some
recent examples \cite{2h-1}. In addition, a light pseudoscalar with
a large $\tan\beta$
 can account for the 3.1$\sigma$ deviation between the SM
predicted and measured values of the muon anomalous magnetic moment
\cite{muon-1,muon-2,muon-3,muon-4}. Due to the experimental
constraints, the type-II 2HDM \cite{type-ii} is very difficult to
explain the muon g-2 anomaly \cite{muon-2,muon-4}, but the
lepton-specific 2HDM (L2HDM) \cite{l2hdm} can still give a valid
explanation \cite{muon-3,muon-4}. Compared to the recent study
\cite{muon-4}, we focus on a light pseudoscalar for which a relative
small $\tan\beta$ is required to account for the muon g-2 anomaly.
For a light pseudoscalar, the 125.5 GeV Higgs decay into the
pseudoscalars is open, and the rare decay $B_s\to \mu^+\mu^-$ can
obtain the additional important contributions from the very light
pseudoscalar exchange diagrams. Therefore, the 125.5 GeV Higgs
signal data and even $B_s\to \mu^+\mu^-$ can give the important
constraints on the very light pseudoscalar. Also we consider the
theoretical constraints, electroweak precision data, the
non-observation of additional Higgs at collider, and the flavor
observables $B\to X_s\gamma$, $\Delta m_{B_s}$ and $\Delta m_{B_d}$.

Our work is organized as follows. In Sec. II we recapitulate the
L2HDM. In Sec. III we introduce the numerical calculations. In Sec.
IV, we show the implications of muon g-2 and experimental data on
the L2HDM. Finally, we give our conclusion in Sec. V.

\section{L2HDM}
The general Higgs potential is written as \cite{2h-poten}
\begin{eqnarray} \label{V2HDM} \mathrm{V} &=& m_{11}^2
(\Phi_1^{\dagger} \Phi_1) + m_{22}^2 (\Phi_2^{\dagger}
\Phi_2) - \left[m_{12}^2 (\Phi_1^{\dagger} \Phi_2 + \rm h.c.)\right]\nonumber \\
&&+ \frac{\lambda_1}{2}  (\Phi_1^{\dagger} \Phi_1)^2 +
\frac{\lambda_2}{2} (\Phi_2^{\dagger} \Phi_2)^2 + \lambda_3
(\Phi_1^{\dagger} \Phi_1)(\Phi_2^{\dagger} \Phi_2) + \lambda_4
(\Phi_1^{\dagger}
\Phi_2)(\Phi_2^{\dagger} \Phi_1) \nonumber \\
&&+ \left[\frac{\lambda_5}{2} (\Phi_1^{\dagger} \Phi_2)^2 + \rm
h.c.\right] + \left[\lambda_6 (\Phi_1^{\dagger} \Phi_1)
(\Phi_1^{\dagger} \Phi_2) + \rm h.c.\right] \nonumber \\
&& + \left[\lambda_7 (\Phi_2^{\dagger} \Phi_2) (\Phi_1^{\dagger}
\Phi_2) + \rm h.c.\right].
\end{eqnarray}
In this paper we focus on the CP-conserving case where all
$\lambda_i$ and $m_{12}^2$ are real. In the L2HDM, a discrete $Z_2$
symmetry is introduced to make $\lambda_6=\lambda_7=0$, and allow
for a soft-breaking term with $m_{12}^2\neq 0$. The two complex
scalar doublets have the hypercharge $Y = 1$,
\begin{equation}
\Phi_1=\left(\begin{array}{c} \phi_1^+ \\
\frac{1}{\sqrt{2}}\,(v_1+\phi_1^0+ia_1)
\end{array}\right)\,, \ \ \
\Phi_2=\left(\begin{array}{c} \phi_2^+ \\
\frac{1}{\sqrt{2}}\,(v_2+\phi_2^0+ia_2)
\end{array}\right).
\end{equation}
Where the vacuum expectation values (VEVs) $v^2 = v^2_1 + v^2_2 =
(246~\rm GeV)^2$, and the ratio of the two VEVs is defined as usual
to be $\tan\beta=v_2 /v_1$. There are five mass eigenstates: two
neutral CP-even $h$ and $H$, one neutral pseudoscalar $A$, and two
charged scalar $H^{\pm}$. We can rotate this basis to the Higgs
basis by a mixing angle $\beta$, where the VEV of $\Phi_2$ field is
zero. In the Higgs basis, the mass eigenstates are obtained from\bea
&&h=\sin(\beta-\alpha)\phi^0_1+\cos(\beta-\alpha)\phi^0_2, \nonumber\\
&&H=\cos(\beta-\alpha)\phi^0_1-\sin(\beta-\alpha)\phi^0_2,
\nonumber\\
&&A=a_2,~~~~~~~~~~H^\pm=\phi^\pm_2. \eea The right fields of the
equations denote the interaction eigenstates in the Higgs basis. The
corresponding masses and couplings of Eq. (\ref{V2HDM}) are changed
in the Higgs basis \cite{higgsbasis}. For example, both $\lambda_6$
and $\lambda_7$ are taken as zero in the physics basis, but the
rotation into the Higgs basis can generate non-zero values for
$\lambda_6$ and $\lambda_7$.

In the Higgs basis, the general Yukawa interactions with no
tree-level FCNC are give \cite{a2hm-1}
\begin{equation} \label{eq:Yukawa1}
 \mathcal{L}_Y = -\frac{\sqrt{2}}{v}\,\Big[M'_d\bar{Q}_L ( \Phi_1 + \kappa_d \Phi_2) d_R
 +M'_u  \bar{Q}_L (\tilde{\Phi}_1 + \kappa_u \tilde{\Phi}_2) u_R
 + M'_\ell\bar{L}_L ( \Phi_1 + \kappa_\ell \Phi_2) \ell_R \Big]
  + \mathrm{h.c.} \,,
\end{equation}
where $\tilde{\Phi}_i(x)=i\tau_2\Phi_i^{\ast}(x)$ and
$M'_{d,u,\ell}$ are the Yukawa matrices.  For the L2HDM, \beq
\kappa_u=\kappa_d=\cot\beta,~~~\kappa_\ell=-\tan\beta.\eeq The
couplings of neutral Higgs bosons with respect to the SM Higgs boson
are give by \bea &&
y^h_{V}=\sin(\beta-\alpha),~~~y^h_f=\sin(\beta-\alpha)+\cos(\beta-\alpha)\kappa_f,\nonumber\\
&&y^H_{V}=\cos(\beta-\alpha),~~~y^H_f=\cos(\beta-\alpha)-\sin(\beta-\alpha)\kappa_f,\nonumber\\
&&y^A_{V}=0,~~~~~~~y^A_u=-i\gamma^5
\kappa_{u},~~~~~~~~y^A_{d,\ell}=i\gamma^5 \kappa_{d,\ell}.\eea Where
$V$ denotes $Z$ and $W$, and $f$ denotes $u$, $d$ and $\ell$. The
charged Higgs couplings are give as
\begin{align} \label{eq:Yukawa2}
 \mathcal{L}_Y & = - \frac{\sqrt{2}}{v}\, H^+\, \Big\{\bar{u} \left[\kappa_d\,V_{CKM} M_d P_R
 - \kappa_u\,M_u V_{CKM} P_L\right] d + \varsigma_\ell\,\bar{\nu} M_\ell P_R \ell
 \Big\}+h.c.,
 \end{align}
where $M_f$ are the diagonal fermion mass matrices.

\section{numerical calculations}
The in-house code is used to calculate the muon g-2, $\chi^2$ fit to
125.5 GeV Higgs signal, $B_s\to \mu^+\mu^-$, $\Delta m_{B_s}$
 and $\Delta m_{B_d}$. $\textsf{2HDMC-1.6.5}$
\cite{2hc-1} is employed to implement the theoretical constraints
from the vacuum stability, unitarity and coupling-constant
perturbativity, and calculate the oblique parameters ($S$, $T$, $U$)
and $\delta\rho$. $\textsf{SuperIso-3.4}$ \cite{spriso} is used to
implement the constraints from  $B\to X_s\gamma$.
$\textsf{HiggsBounds-4.1.3}$ \cite{hb} is employed to implement the
exclusion constraints from the neutral and charged Higgses searches
at the LEP, Tevatron and LHC at 95\% confidence level. Now we
introduce the calculations of some constraints, which are the main
motivations of this paper:

$\textsf{Muon g-2:}$ The recent measurement on the muon anomalous
magnetic moment is $a_\mu^{exp}=(116592091\pm63)\times10^{-11}$
\cite{muonexp}, which has approximately 3.1$\sigma$ deviation from
the SM prediction \cite{muonsm,muonliu}, $\Delta
a_\mu=a_\mu^{exp}-a_\mu^{SM}=(262\pm85)\times10^{-11}$
\cite{muon-4}.

In the L2HDM, $a_\mu$ gets the additional contributions from the
one-loop diagrams induced by the Higgs bosons and also from the
two-loop Barr-Zee diagrams mediated by $A$, $h$ and $H$. For the
one-loop contributions \cite{muon1loop},
 \beq
    \Delta a_\mu^{\mbox{$\scriptscriptstyle{\rm 2HDM}$}}({\rm 1loop}) =
    \frac{G_F \, m_{\mu}^2}{4 \pi^2 \sqrt{2}} \, \sum_j
    \left (y_{\mu}^j \right)^2  r_{\mu}^j \, f_j(r_{\mu}^j),
\label{amuoneloop}
\end{equation}
where $j = h,~ H,~ A ,~ H^\pm$, $r_{\mu}^ j =  m_\mu^2/M_j^2$. For
$r_{\mu}^j\ll$ 1, \beq
    f_{h,H}(r) \simeq- \ln r - 7/6,~~
    f_A (r) \simeq \ln r +11/6, ~~
    f_{H^\pm} (r) \simeq -1/6.
    \label{oneloopintegralsapprox3}
\eeq

For two-loop contributions, \beq
    \Delta a_\mu^{\mbox{$\scriptscriptstyle{\rm 2HDM}$}}({\rm 2loop-BZ})
    = \frac{G_F \, m_{\mu}^2}{4 \pi^2 \sqrt{2}} \, \frac{\alpha_{\rm em}}{\pi}
    \, \sum_{i,f}  N^c_f  \, Q_f^2  \,  y_{\mu}^i  \, y_{f}^i \,  r_{f}^i \,  g_i(r_{f}^i),
\label{barr-zee}
\end{equation}
where $i = h,~ H,~ A$. $m_f$, $Q_f$ and $N^c_f$ are the mass,
electric charge and number of color degrees of freedom of the
fermion $f$ in the loop. The functions $g_i(r)$ are
\cite{muon-1,muon-2}\beq
    g_{h,H}(r) = \int_0^1 \! dx \, \frac{2x (1-x)-1}{x(1-x)-r} \ln
    \frac{x(1-x)}{r}, ~~~g_{A}(r) = \int_0^1 \! dx \, \frac{1}{x(1-x)-r} \ln
    \frac{x(1-x)}{r}.
\eeq The contributions of the CP-even (CP-odd) Higgses to $a_\mu$
are negative (positive) at the two-loop level and positive
(negative) at one-loop level. As $m^2_f/m^2_\mu$ could easily
overcome the loop suppression factor $\alpha/\pi$, the two-loop
contributions may be larger than one-loop ones. In the L2HDM, since
the CP-odd Higgs coupling to the tau lepton is proportional to
$\tan\beta$, the L2HDM can enhance sizably the muon g-2 for a light
CP-odd Higgs with a large $\tan\beta$.

$\textsf{Global fit to the 125.5 GeV Higgs signal data:}$ We take
the light CP-even Higgs as the 125.5 GeV Higgs. Using the method
taken in \cite{chi}, we perform a global fit to the 125.5 GeV Higgs
data of 29 channels after ICHEP 2014, which are summarized in the
Tables I-V of \cite{kmdata}. A number of new measurements or updates
of existing ones were presented by ATLAS and CMS Collaborations
\cite{k-7,k-8,atl-1,atl-2,atl-3,atl-4,cms-1,cms-2,cms-3,tev-1}. The
signal strength for the $i$ channel is defined as
\beq\mu_i=\epsilon_{ggh}^i R_{ggH}+\epsilon_{VBF}^i
R_{VBF}+\epsilon_{VH}^i R_{VH}+\epsilon_{t\bar{t}H}^i
R_{t\bar{t}H}.\eeq Where $R_{j}=\frac{(\sigma \times
BR)_j}{(\sigma\times BR)_j^{SM}}$ with $j$ denoting the partonic
processes $ggH,~VBF,~VH,$ and $t\bar{t}H$. $\epsilon_{j}^i$ denotes
the assumed signal composition of the partonic process $j$, which
are given in Tables I-V of \cite{kmdata}. For an uncorrelated
observable $i$, \beq
\chi^2_i=\frac{(\mu_i-\mu^{exp}_i)^2}{\sigma_i^2},\eeq where
$\mu^{exp}_i$ and $\sigma_i$ denote the experimental central value
and uncertainty for the $i$ channel. We retain the uncertainty
asymmetry in our calculations. For the two correlated observables,
we take \beq \chi^2_{i,j}=\frac{1}{1-\rho^2}
\left[\frac{(\mu_i-\mu^{exp}_i)^2}{\sigma_i^2}+\frac{(\mu_j-\mu^{exp}_j)^2}{\sigma_j^2}
-2\rho\frac{(\mu_i-\mu^{exp}_i)}{\sigma_i}\frac{(\mu_j-\mu^{exp}_j)}{\sigma_j}\right],\eeq
where $\rho$ is the correlation coefficient. We sum over $\chi^2$ of
the 29 channels, and pay particular attention to the surviving
samples with $\chi^2-\chi^2_{\rm min} \leq 6.18$, where $\chi^2_{\rm
min}$ denotes the minimum of $\chi^2$. These samples correspond to
the 95.4\% confidence level regions in any two dimensional plane of
the model parameters when explaining the Higgs data (corresponding
to be within $2\sigma$ range).

$\textsf{$B_s\to \mu^+\mu^-$:}$ The LHCb \cite{lhcb-bsuu} and CMS
\cite{cms-bsuu} measurements lead to the weighted average,
$\bar{B}(B_s\to\mu^+\mu^-)_{exp}=(2.9\pm0.7)\times10^{-9}$
\cite{aver-bsuu}, which is well agreement with the latest SM
prediction,
$\bar{B}(B_s\to\mu^+\mu^-)_{SM}=(3.65\pm0.23)\times10^{-9}$
\cite{sm-bsuu}. Recently, Ref. \cite{li-bsuu} presents a complete
one-loop calculation of the contributions of Aligned 2HDM to $B_s\to
\mu^+\mu^-$. Following the method taken in Ref. \cite{li-bsuu}, we
define
\begin{equation} \label{rbsuu}
 \overline{R}_{s\mu}  \,\equiv\,  \frac{\overline{\mathcal{B}}(B_s\to \mu^+\mu^-)}
 {\overline{\mathcal{B}}(B_s\to \mu^+\mu^-)_{\rm SM}}
\, =\,
\bigg[\,|P|^2+\Big(1-\frac{\Delta\Gamma_s}{\Gamma^s_L}\,\Big)|S|^2\bigg]\,,
\end{equation}
where the CKM matrix elements and hadronic factors cancel out.
Combining the SM prediction with the experimental result,
$\bar{R}_{s\mu}=0.79\pm 0.2$ is required.
\begin{align}
 P &\equiv\, \frac{C_{10}}{C^{\rm SM}_{10}} + \frac{M^2_{B_s}}{2M^2_W}
 \left(\frac{m_b}{m_b+m_s}\right)\,\frac{C_P-C_P^{\mathrm{SM}}}{C^{\rm
 SM}_{10}},
  \,
 \label{eq:P}\\[0.2cm]
 S &\equiv\, \sqrt{1-\frac{4m^2_\mu}{M^2_{B_s}}}\; \frac{M^2_{B_s}}{2M^2_W}
 \left(\frac{m_b}{m_b+m_s}\right)\,\frac{C_S-C_S^{\mathrm{SM}}}{C^{\rm SM}_{10}}
 \,.
 \label{eq:S}
\end{align}
The 2HDM can give the additional contributions to coefficient
$C_{10}$ by the $Z$-penguin diagrams with the charged Higgs loop.
Unless there are large enhancements for $C_P$ and $C_S$, their
contributions can be neglected due to the suppression of the factor
$M_{B_s}^2/M_W^2$. For example, the $C_P$ and $C_S$ of type-II 2HDM
can be dominant due to the enhancement of the large $\tan^2\beta$
terms \cite{tb2}. Although such large $\tan^2\beta$ terms are absent
in the L2HDM, $C_P$ can obtain the important contributions from the
CP-odd Higgs exchange diagrams for $m_A$ is very small. Such
contributions are also sensitive to $m_{H^\pm}$ and small
$\tan\beta$. For the large $\tan\beta$, the terms proportional to
$\cot\beta$ and the higher order terms can be neglected.

Using the formulas in \cite{li-bsuu}, we calculate the parameter $P$
and $S$ in the L2HDM. Note that the mixing of two CP-even Higgses in
this paper is different from \cite{li-bsuu}, therefore some
corresponding couplings need be replaced.

In our calculations, $m_h=$ 125.5 GeV is fixed, and the input
parameters are  $\sin(\beta-\alpha)$, $\tan\beta$, the physical
Higgs masses ($m_H$, $m_A$, $m_{H^{\pm}}$) and the coupling of $hAA$
($\lambda_{hAA}$), where $\lambda_{hAA}$ is used to replace the
soft-breaking parameter $m_{12}^2$. We focus on 5 GeV $<m_A<$ 62.75
GeV, and such light CP-odd Higgs will be more strongly constrained,
especially for the 125.5 GeV Higgs signal. Assuming that the Higgs
couplings to fermions and gauge bosons are the same as the SM,
$Br(h\to AA)$ is larger than $40\%$ for $|\lambda_{hAA}|>$ 20 GeV
and $m_A <62.5$ GeV. Therefore, we scan $\lambda_{hAA}$ in the range
of -20 GeV $\sim$ 20 GeV. In addition to that the theoretical
constraints are satisfied, we require the L2HDM to explain the
experimental data within the 2$\sigma$ range, and fit the current
Higgs signal data at the $2\sigma$ level. The experimental values of
electroweak precision data, $B\to X_s\gamma$, $\Delta m_{B_s}$ and
$\Delta m_{B_d}$ are taken from \cite{pdg2014}.

\section{results and discussions}
Without the constraint of muon g-2, we find a surviving sample with
a minimal value of $\chi^2$ fit to the Higgs signal data,
$\chi^2_{min}\simeq16.95$, which is slightly smaller than SM value ,
17.0. The corresponding input parameters are, \bea&&
\sin(\beta-\alpha)\simeq-0.999994,~\tan\beta\simeq5.16,~m_h=125.5
~{\rm{GeV}},~m_H\simeq130.35~{\rm{GeV}},\nonumber\\
&&m_A=61.50~{\rm{GeV}},~m_{H^\pm}=146.21~{\rm{GeV}},\nonumber\\
&&\lambda_{hAA}\simeq-0.47~{\rm{GeV}}~
(m_{12}^2=2174.84~\rm{GeV^2}).~ \eea Our numerical results show that
for the surviving samples within the 2$\sigma$ range of $\chi^2$,
$Br(h\to AA)$ is required to be smaller than 24\%. Considering the
constraint of muon g-2, the minimal value is
$\chi^2_{min}\simeq17.21$ and the input parameters are \bea&&
\sin(\beta-\alpha)\simeq0.999712,~\tan\beta\simeq84.90,~m_h=125.5
~{\rm{GeV}},~m_H\simeq504.34~{\rm{GeV}},\nonumber\\
&&m_A=57.00~{\rm{GeV}},~m_{H^\pm}=509.18~{\rm{GeV}},\nonumber\\
&&\lambda_{hAA}\simeq-0.95~{\rm{GeV}}~
(m_{12}^2=2995.28~\rm{GeV^2}).~ \eea

\begin{figure}[tb]
 \epsfig{file=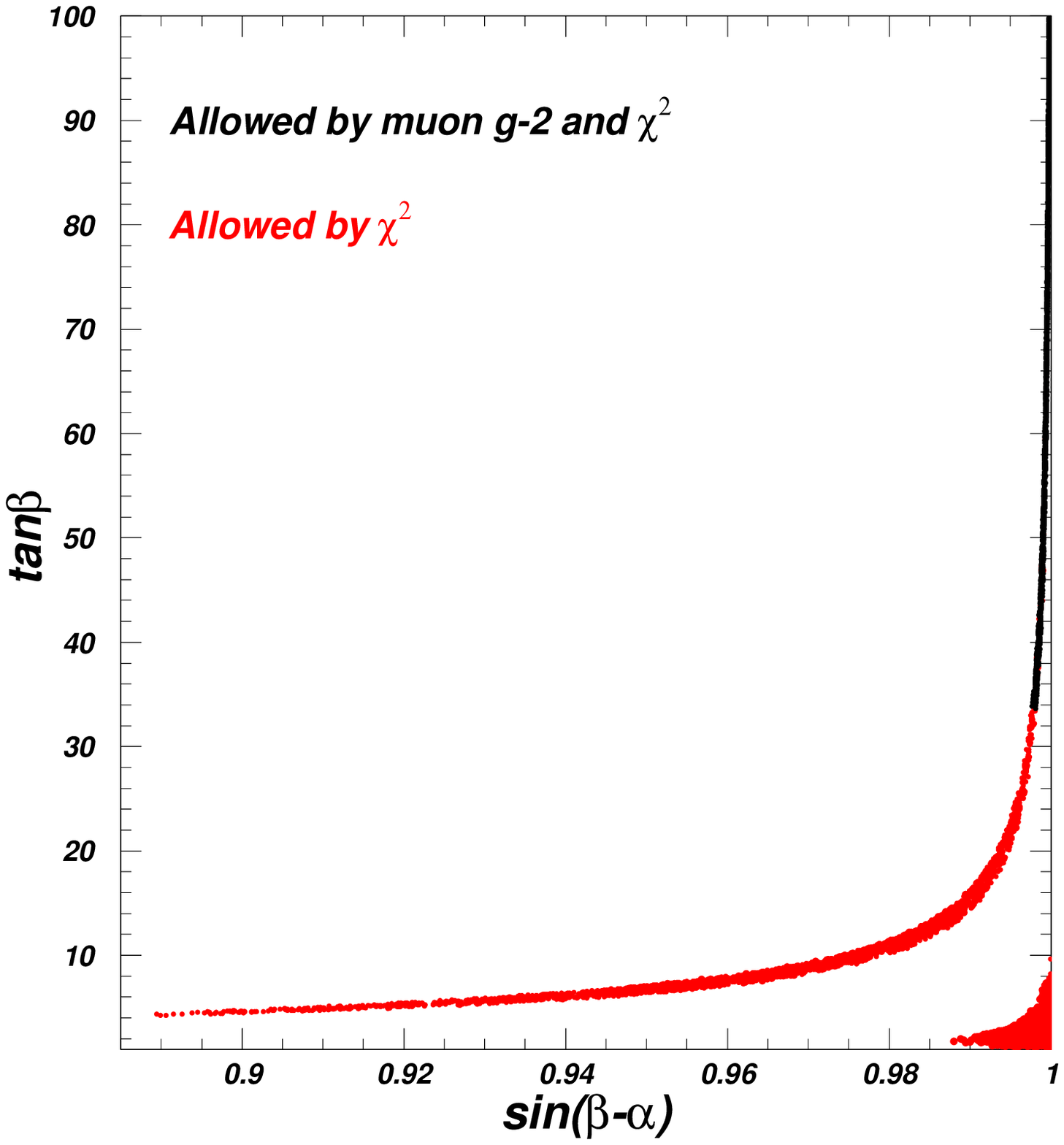,height=7.5cm}
  \epsfig{file=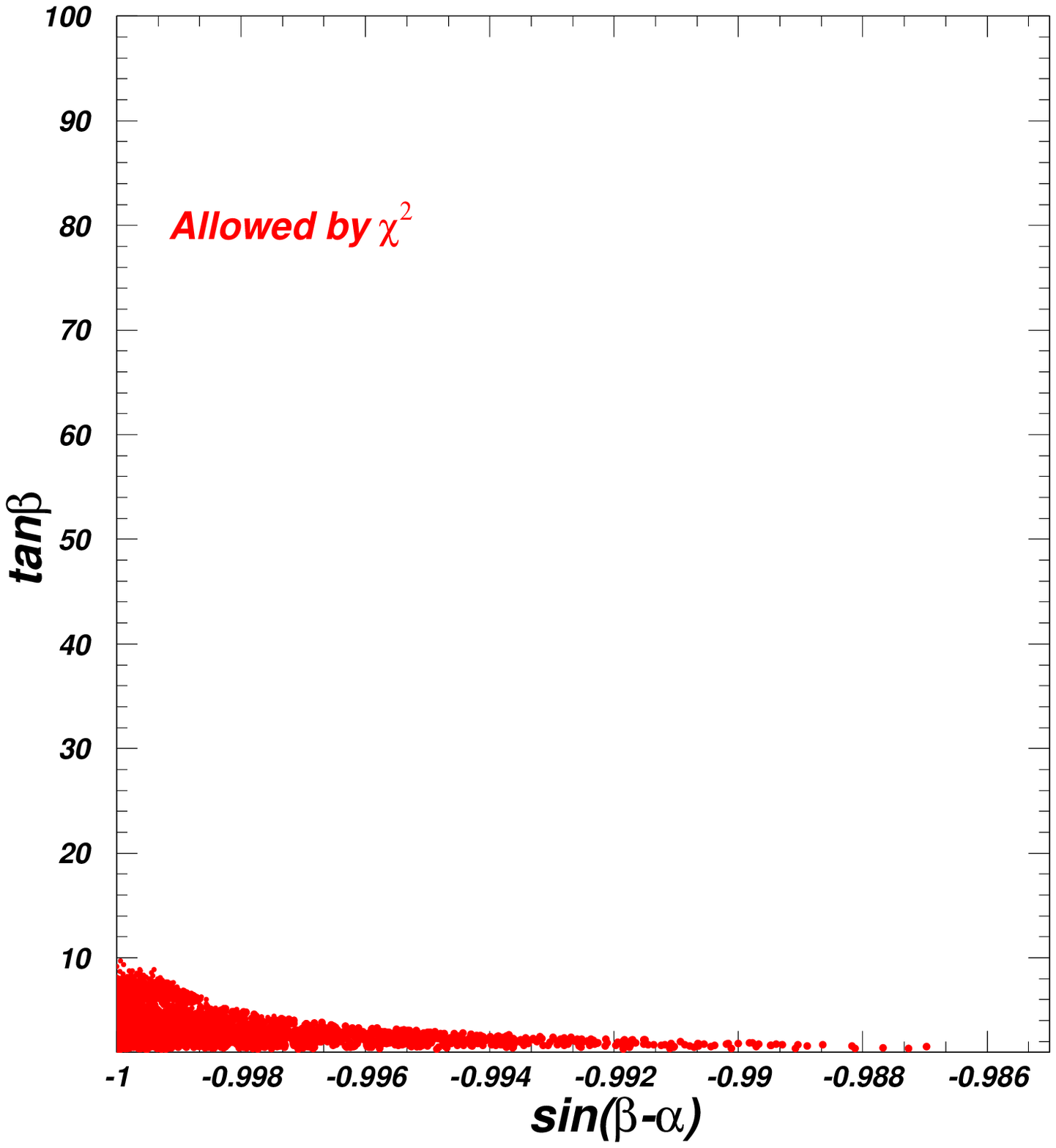,height=7.5cm}
\vspace{-0.3cm} \caption{The scatter plots of surviving samples
projected on the plane of $\sin(\beta-\alpha)$ versus $\tan\beta$.}
\label{sbatb}
\end{figure}

In Fig. \ref{sbatb}, we project the surviving samples on the plane
of $\sin(\beta-\alpha)$ versus $\tan\beta$. Without the constraint
of muon g-2, the surviving samples lie in two different regions. In
one region, the 125.5 GeV Higgs couplings are near the SM values,
called the SM-like region. In the other region, the Higgs couplings
to leptons have opposite sign to the corresponding couplings to
$VV$, called the wrong-sign Yukawa coupling region. In the SM-like
region, the absolute value of $\sin(\beta-\alpha)$ is required to be
larger than 0.986, while $\sin(\beta-\alpha)$ is allowed to have
more sizable deviation from 1.0 in the wrong-sign Yukawa coupling
region, $\sin(\beta-\alpha)>$  0.89. This can be understandable from
the Higgs couplings to leptons. For the wrong-sign Yukawa coupling
and SM-like Yukawa coupling, the Higgs couplings to lepton are
respectively
\beq\sin(\beta-\alpha)-\tan\beta\cos(\beta-\alpha)=-1+\varepsilon,~~
\sin(\beta-\alpha)-\tan\beta\cos(\beta-\alpha)=1-\varepsilon,
\label{hlleq}\eeq where the absolute value of $\varepsilon$ is much
smaller than 1.0. For $\sin(\beta-\alpha)$ approaches to 1.0,
$\cos(\beta-\alpha)$ of the former is much larger than that of the
latter for the same $\tan\beta$.

\begin{figure}[tb]
 \epsfig{file=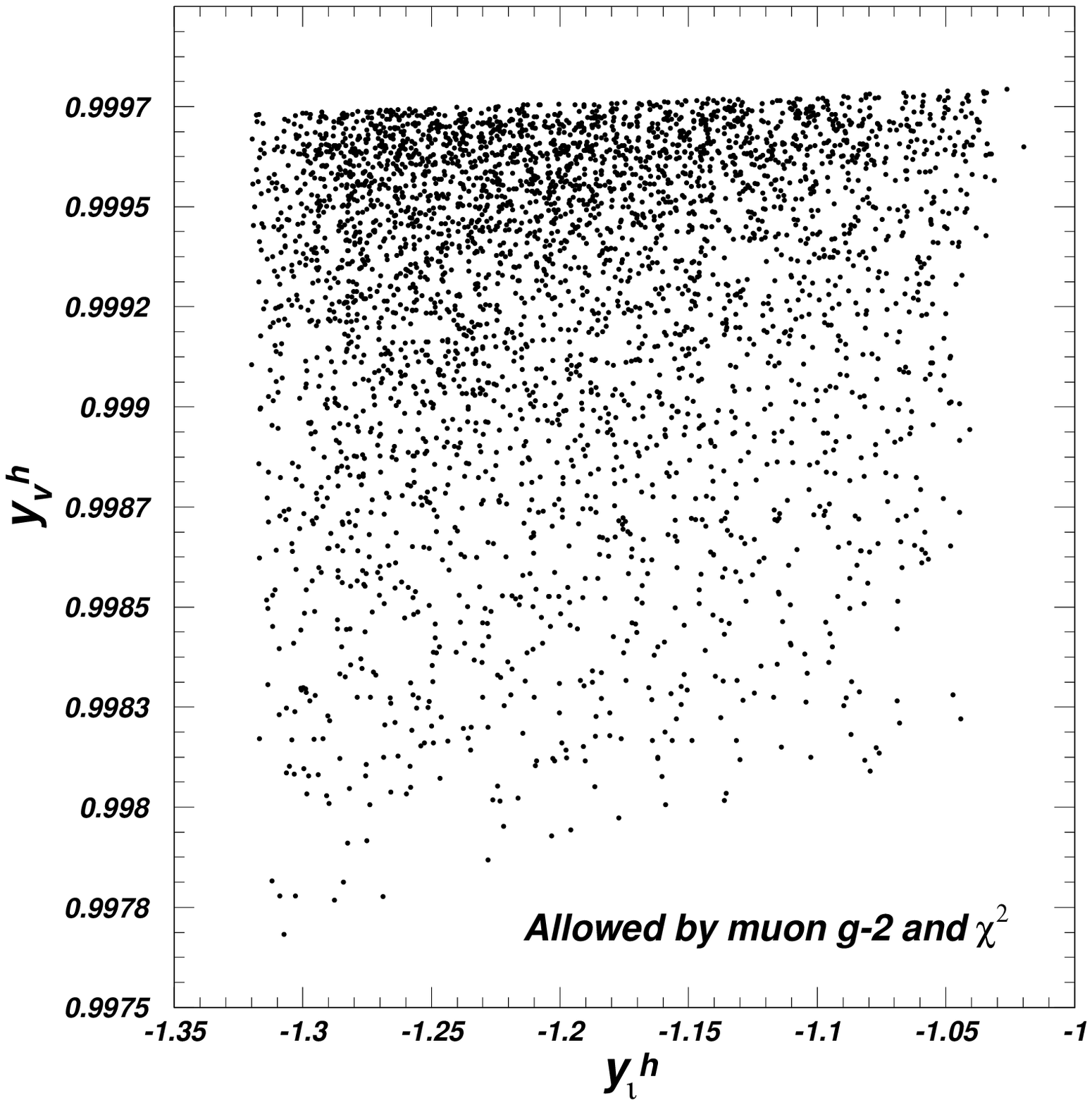,height=7.5cm}
  \epsfig{file=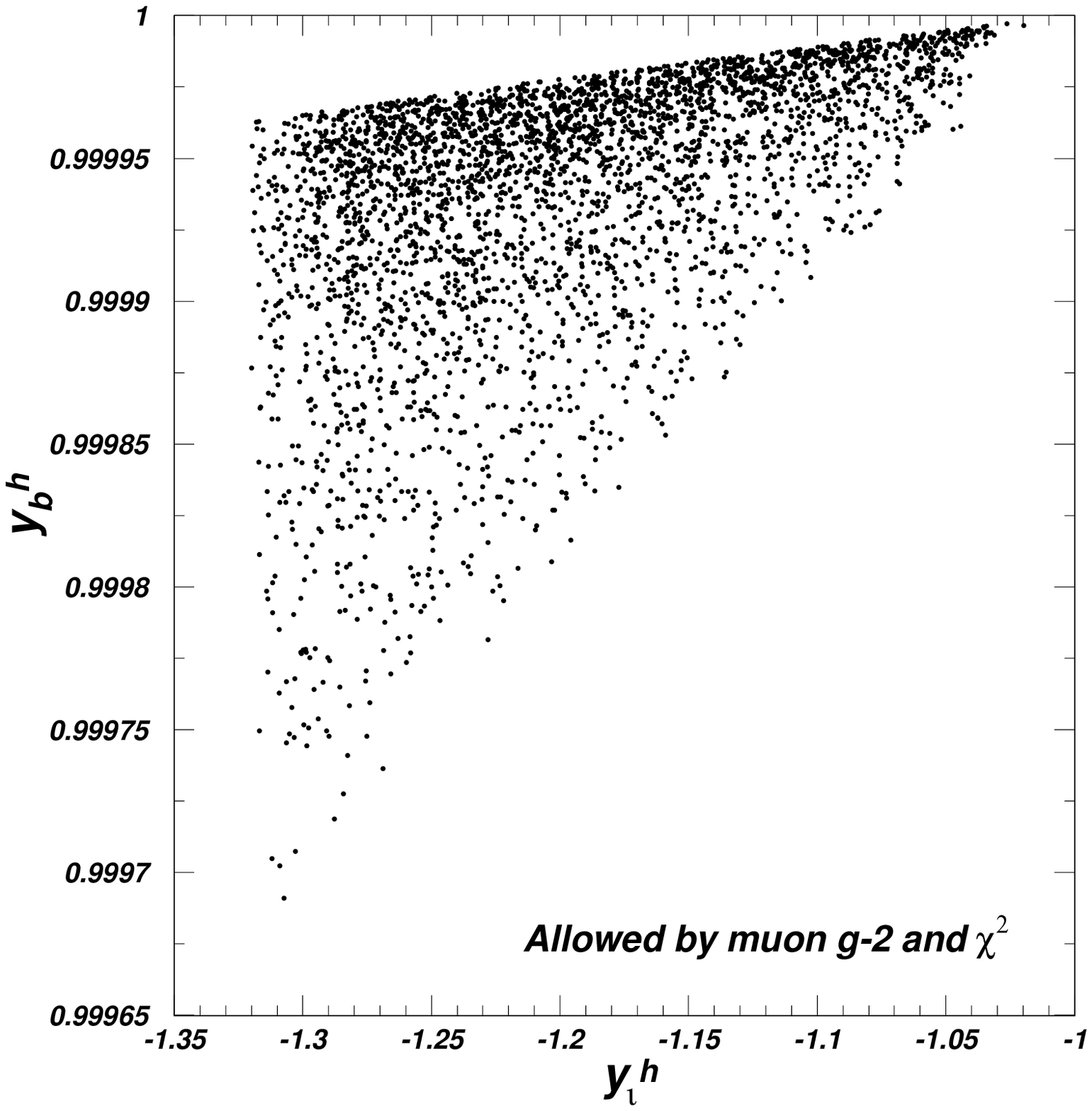,height=7.5cm}
\vspace{-0.3cm} \caption{The scatter plots of surviving samples with
the 2$\sigma$ ranges of muon g-2 and $\chi^2$ projected on the
planes of the coupling $h\ell\bar{\ell}$ versus $hVV$, and
$h\ell\bar{\ell}$ versus $hb\bar{b}$, respectively.} \label{hcoup}
\end{figure}

Including the constraint of muon g-2, the surviving samples favor
the wrong-sign Yukawa coupling region. The corresponding
$\sin(\beta-\alpha)$ approaches to 1.0 as increasing of $\tan\beta$,
leading a small $\cos(\beta-\alpha)$ which ensures the absolute
value of the coupling $h\ell\bar{\ell}$ around SM value. To account
for the muon g-2, L2HDM has to provide a very large pseudoscalar
coupling to lepton, and $\tan\beta$ is required to be larger than 34
as shown in the left panel of Fig. \ref{sbatb}. For such large
$\tan\beta$, Eq. (\ref{hlleq}) shows that $y_\ell^h$ is much smaller
than $-1.0$ for $\sin(\beta-\alpha)$ approaches to -1.0. As a
result, the absolute value of the 125.5 GeV Higgs couplings to
leptons have the sizable deviations from SM predictions, which is
excluded by the 2 $\sigma$ range of $\chi^2$. In addition, according
to Eq. (\ref{hlleq}), such large $\tan\beta$ leads to $y_\ell^h <$ 0
although $\sin(\beta-\alpha)$ approaches to 1.0. Therefore, the
surviving samples lie in the wrong-sign Yukawa coupling region.

 In Fig.
\ref{hcoup}, the surviving samples are projected on the planes of
Higgs couplings. The Higgs couplings to $VV$ and quarks are very
closed to SM values, but the Higgs couplings to leptons have the
opposite sign to the SM values, and over $30\%$ deviation from the
SM values.

\begin{figure}[tb]
 \epsfig{file=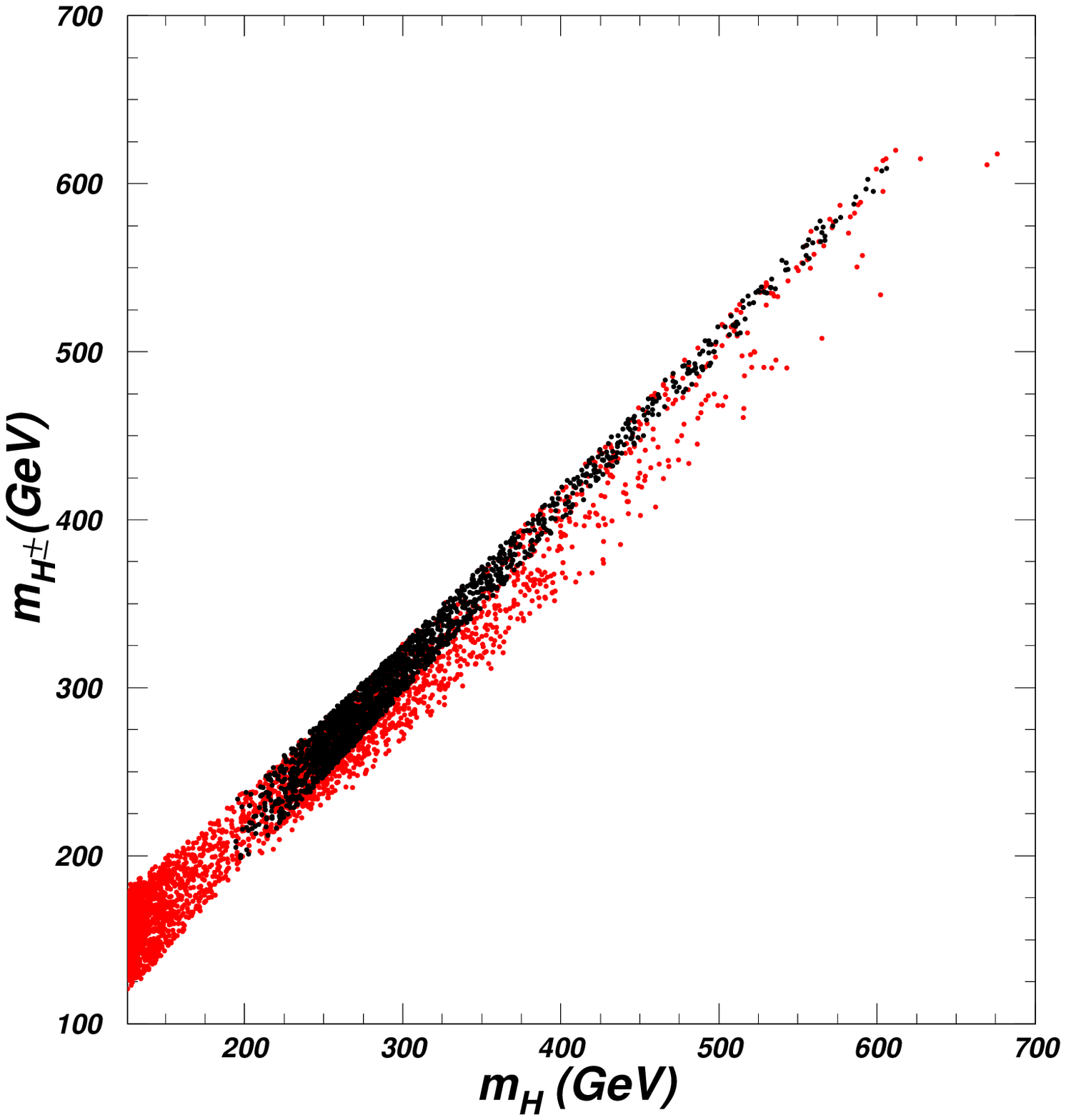,height=7.5cm}
  \epsfig{file=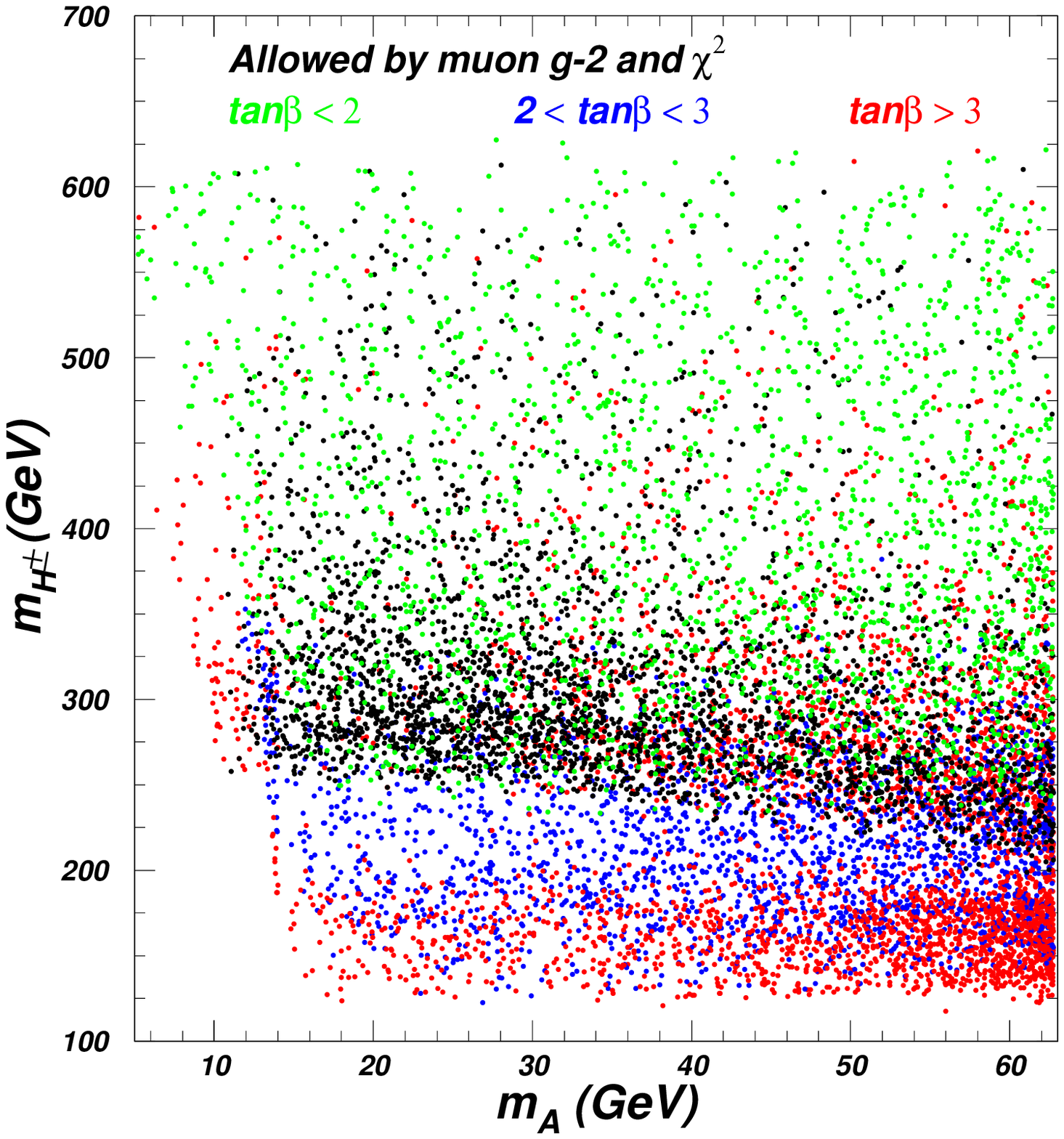,height=7.5cm}
\vspace{-0.3cm} \caption{Left panel: Same as Fig. \ref{sbatb}, but
projected on the plane of $m_H$ versus $m_{H^\pm}$. Right panel: The
scatter plots of surviving samples within the 2$\sigma$ range of
$\chi^2$ projected on the plane of $m_A$ versus $m_{H^\pm}$.}
\label{higgmass}
\end{figure}

In Fig. \ref{higgmass}, the surviving samples are projected on the
planes of $m_H$ versus $m_{H^\pm}$ and $m_A$ versus $m_{H^\pm}$,
respectively. The left panel shows that there is a small mass
difference between $m_H$ and $m_{H^\pm}$ for the surviving samples,
especially for including the constraint of muon g-2. Such small mass
difference is mainly required by the electroweak parameter $\rho$ to
produce a pseudoscalar with mass in the range of 5 GeV $\sim$ 62.75
GeV. As shown in Fig. \ref{sbatb}, the experimental data of muon g-2
require $\tan\beta > $ 34. For such large $\tan\beta$, the charged
Higgs has a very large coupling to lepton, and the search
experiments of charged Higgs give the lower bound of the charged
Higgs mass, $m_{H^\pm}> 200$ GeV. Due to the small mass difference
between $m_H$ and $m_{H^\pm}$ induced by the parameter $\rho$, $m_H$
is required to be larger than 200 GeV. In Ref. \cite{muon-4}, the
authors took the limiting case of $\beta-\alpha$= $\frac{\pi}{2}$
and several fixed values of $m_H-m_{H^{\pm}}$ and $\lambda_1$, and
found that the theoretical constraints and electroweak precision
data give the upper bound of charged Higgs, $m_{H^{\pm}}\leq$ 200
GeV for $m_A<$ 100 GeV. In this paper, we scan the whole parameter
space, and find that $m_{H^{\pm}}$ is allowed to be as high as 600
GeV. Our results are consistent with those of many other papers,
such as Ref. \cite{mhplus1}, Ref. \cite{mhplus2} and Ref.
\cite{muon-3}.

The right panel shows that $\tan\beta$ is required to be larger than
2.0 for $m_{H^\pm}<$ 230 GeV, and the main constraints are from the
flavor observables $\Delta m_{B_s}$ and $\Delta m_{B_d}$. In
addition, for $m_A <$ 16 GeV, there is strong correlation between
$m_A$ and $m_{H^\pm}$ due to the constraint of $B_s\to\mu^+\mu^-$.
In particular, $m_{H^\pm}$ is required to be larger than 450 GeV for
$\tan\beta < $ 2 and $m_A<$ 10 GeV, and $m_{H^\pm}$ is allowed to be
sharply decreased with the increasing of $m_A$. For $m_A >$ 20 GeV,
the contributions from the exchange of $A$ diagrams to the
coefficient $C_P$ are difficult to overcome the suppression factor
$M_{B_s}^2/M_W^2$, therefore $B_s\to\mu^+\mu^-$ is not sensitive to
$m_A$. Also the constraint of $B_s\to \mu^+\mu^-$ on $m_A$ and
$m_{H^\pm}$ can be relaxed by a modest large $\tan\beta$, but not
sensitive to the enough large $\tan\beta$. Including the constraint
of muon g-2, $m_A$ is allowed to be as low as 10 GeV, but the
corresponding $m_{H^\pm}$ is required to be larger than 250 GeV.

\begin{figure}[tb]
 \epsfig{file=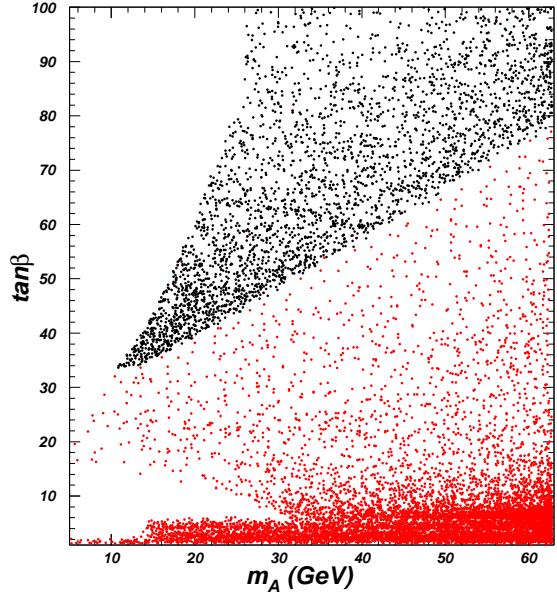,height=8.0cm}
\vspace{-0.3cm} \caption{Same as Fig. \ref{sbatb}, but projected on
the plane of $m_A$ versus $\tan\beta$.} \label{matb}
\end{figure}

In Fig. \ref{matb}, the surviving samples are projected on the plane
of $m_A$ versus $\tan\beta$. For $m_A<$ 26 GeV, the upper bound of
$\tan\beta$ is strongly constrained by the exclusion experiments of
Higgs at the collider, and some intermediate values are excluded by
the 2$\sigma$ constraint of $\chi^2$ fit to the Higgs signal.
Including the constraint of muon g-2, the range of $\tan\beta$ is
sizably narrowed with $m_A$ approaching to 10 GeV.

\section{Conclusion}
In the L2HDM, a light pseudoscalar with a large $\tan\beta$ can
account for the muon g-2 anomaly. Assuming that the light CP-even
Higgs is the 125.5 GeV Higgs, we study the implications of the
relevant theoretical and experimental constraints on a pseudoscalar
with the mass below the half of 125.5 GeV, especially for the muon
g-2 anomaly, 125.5 GeV Higgs signal and $B_s\to \mu^+\mu^-$. We find
that the pseudoscalar can be allowed to be as low as 10 GeV, and
$\tan\beta$ is required to be larger than 34. As the increasing of
$\tan\beta$, $\sin(\beta-\alpha)$ is very closed to 1.0. For $m_A$
approaches to 10 GeV, the range of $\tan\beta$ is sizably narrowed,
and $m_{H^\pm}$ is required to be larger than 250 GeV. In addition,
the 125.5 GeV Higgs couplings to leptons are favored to have
opposite sign to the couplings to gauge bosons and quarks, and over
30\% deviation from the SM values.

\section*{Acknowledgment}
We would like to thank Xin-Qiang Li for helpful discussions. This
work was supported by the National Natural Science Foundation of
China (NNSFC) under grant No. 11105116.

\end{document}